\documentclass[12pt]{article}
\usepackage{epsfig}
\font\tf=cmr7 scaled\magstep 1
\makeatletter
\@addtoreset{equation}{on}

\makeatother
\begin{document}
\begin{flushright}
{\tt hep-th/0512234} \\
KEK-TH-1037\\
HIP-2005-57/TH\\ 
\end{flushright}
\vspace{2mm}
\begin{center}
{\large \bf 
 Color Superconductivity in ${\cal N}=2$ 
 Supersymmetric Gauge Theories}
\end{center}
\vspace{8mm}
\begin{center}
\normalsize
{\large \bf  Masato Arai $^a$
\footnote{E-mail: masato.arai@helsinki.fi}
and Nobuchika Okada $^{b,c}$
\footnote{E-mail: okadan@theory.kek.jp}  
}
\end{center}
\vskip 1.2em
\begin{center}
{\it 
$^a$High Energy Physics Division, 
             Department of Physical Sciences,
             University of Helsinki \\
 and Helsinki Institute of Physics,
 P.O.Box 64, FIN-00014, Finland \\
\vskip 1.0em
$^b$Theory Group, KEK, Tsukuba 305-0801, Japan \\
$^c$ Department of Particle and Nuclear Physics,  
The Graduate University \\
for Advanced Studies (Sokendai), 
Oho 1-1, Tsukuba 305-0801, Japan 
}
\end{center}
\vskip 1.0cm
\begin{center}
{\large Abstract}
\vskip 0.7cm
\begin{minipage}[t]{14cm}
\baselineskip=19pt
\hskip4mm
We study vacuum structure of ${\cal N}=2$ supersymmetric (SUSY) QCD, 
 based on the gauge group $SU(2)$ 
 with $N_f=2$ flavors of massive hypermultiplet quarks, 
 in the presence of non-zero baryon chemical potential ($\mu$). 
The theory has a classical vacuum preserving baryon number symmetry, 
 when a mass term, which breaks ${\cal N}=2$ SUSY 
 but preserves ${\cal N}=1$ SUSY, 
 for the adjoint gauge chiral multiplet ($m_{\mbox{\tf ad}}$) 
 is introduced. 
By using the exact result of ${\cal N}=2$ SUSY QCD, 
 we analyze low energy effective potential 
 at the leading order of perturbation 
 with respect to small SUSY breaking parameters, 
 $\mu$ and $m_{\mbox{\tf ad}}$. 
We find that the baryon number is broken 
 as a consequence of the $SU(2)$ strong gauge dynamics, 
 so that color superconductivity dynamically takes place 
 at the non-SUSY vacuum. 
\end{minipage}
\end{center}
\newpage
%
%
\def\barr{\begin{eqnarray}}
\def\earr{\end{eqnarray}}

\section{Introduction}

Recently phase structure of quark matter 
 under non-zero baryon density has been intensively studied.
In particular, it has been emphasized that quark matter is expected 
 to be in a BCS-type superconducting phase 
 at high baryon densities \cite{wilczek}.
In the case with high baryon densities 
 the Fermi surface lies at a high energy regime, 
 and perturbative description of QCD is applicable. 
There exists an attractive interaction between quarks 
 through the exchange of one gluon in a color anti-symmetric 
 $\bar{{\bf 3}}$ state.
Therefore, QCD at high densities is expected to behave 
 as a color superconductor. 
One of important consequences of color superconductivity 
 is spontaneous breaking of baryon number symmetry. 
In fact, spontaneous breaking of the baryon number symmetry 
 is observed in high density QCD for three or more quark flavors 
 with breaking of color and flavor symmetries 
 (color flavor locking) \cite{wilczek2}. 
However, in the case of low baryon densities, 
 where QCD is in strong coupling regime, 
 perturbative calculations is no longer applicable. 
It is an open question
 whether the color superconducting phase appears 
 even at low baryon densities. 

If some QCD-like theories exist 
 which is calculable beyond perturbation, 
 it would be interesting to investigate vacuum structure 
 of such theories in the presence of non-zero baryon 
 chemical potential. 
There have been prominent developments of SUSY gauge theory 
 in the past decade. 
In ${\cal N}=1$ SUSY gauge theory, the exact low energy effective 
 superpotential was derived by using holomorphy properties 
 of the superpotential and the gauge kinetic function 
 (for review, see Ref. \cite{seiberg}). 
Seiberg and Witten derived the exact low energy Wilsonian effective action 
 for ${\cal N}=2$ SUSY $SU(2)$ Yang-Mills theory and generalized it 
 to the case up to four massive hypermultiplets by using holomorphy 
 and duality arguments \cite{witten}. 
By using the exact superpotential, 
 Harnik et al. \cite{murayama} studied vacuum structure 
 of ${\cal N}=1$ SUSY QCD with non-zero baryon chemical potential. 
They found that the baryon number symmetry is spontaneously broken 
 due to the effect of the strong gauge dynamics 
 and also examined the breaking pattern of flavor symmetries. 
However, in ${\cal N}=1$ SUSY framework, 
 the region that we can investigate is limited, 
 because effective K{\"a}hler potential, 
 which is not a holomorphic variable, 
 can not be determined beyond perturbations, 
 while effective superpotential is exact. 

In this paper, we study vacuum structure of  
 ${\cal N}=2$ SUSY QCD in the presence of non-zero baryon 
 chemical potential. 
Remarkable point is that in the theory 
 we can derive exact low energy effective theory 
 including effective K{\"a}hler potential. 
The theory we will investigate is based on  
 the gauge group $SU(2)$ with $N_f=2$ flavors 
 of hypermultiplet quarks with the same masses. 
In order to provide definite classical vacua, we also introduce 
 ${\cal N}=1$ SUSY preserving mass term for the gauge adjoint 
 chiral multiplet. 
At the classical level, the theory has two discrete vacua 
 in some parameter region for the baryon chemical potential $\mu$, 
 the common masses of hypermultiplet quarks $m$ 
 and the adjoint mass $m_{\mbox{\tf ad}}$. 
One is a local minimum where baryon number symmetry 
 is preserved, the other is a global one breaking the baryon number. 
Our main interest is whether the baryon number symmetry 
 is spontaneously broken or not after taking all quantum corrections 
 into account, and so we concentrate on the classical vacuum preserving 
 the baryon number symmetry. 
By using the exact results of ${\cal N}=2$ SUSY QCD, 
 we derive low energy effective theory and 
 investigate how the classical vacuum is deformed 
 due to the $SU(2)$ strong gauge dynamics. 

As the same as in the analysis of ${\cal N}=1$ SUSY theory 
 \cite{murayama}, 
 there is a clear limitation on our analysis, 
 namely SUSY breaking parameters in our theory, 
 the chemical potential $\mu$ and the adjoint mass $m_{\mbox{\tf ad}}$, 
 should be much smaller than the dynamical scale of 
 the $SU(2)$ gauge interaction $\Lambda$. 
Keeping them in such a region, 
 we use the exact results of ${\cal N}=2$ SUSY QCD 
 and derive low energy effective theory 
 at the leading order of perturbation 
 with respect to such small SUSY breaking parameters. 
Note that the parameter $\mu$ we consider is small, 
 where gauge coupling is in strong coupling regime. 
Thus, the results of our analysis would give 
 some insights for color superconducting phase 
 in strong coupling regime, 
 complementary to the results obtained in perturbative 
 analysis of non-SUSY QCD with large baryon chemical potential. 
Unfortunately, in our analysis, diquark fields do not appear 
 as dynamical variables and 
 we cannot examine their condensations $\langle qq \rangle$ 
 (for a related work, see Ref. \cite{MT}). 

The organization of this paper is as follows. 
In section 2, we define our classical Lagrangian 
 and find classical vacua. 
In Section 3, we derive the low energy effective action 
 by using exact results of ${\cal N}=2$ SUSY QCD. 
Potential analysis is performed numerically in section 4.
Finally we summarize our results.

\section{Vacuum structure of classical theory}
In this section, we first define our classical Lagrangian 
 and analyze its classical vacuum. 
We will find that the theory has at least one vacuum 
 preserving the baryon number symmetry. 

Our theory is based on the ${\cal N}=2$ $SU(2)$ gauge theory 
 with $N_f=2$ quark hypermultiplets having the same masses $m$. 
In terms of ${\cal N}=1$ superfield language, 
 this theory is described by 
 two pairs of chiral quark and anti-quark superfields 
 $Q_i$ and $\tilde{Q}_i^\dagger$ ($i=1,2$ denotes the flavor index) 
 in the (anti) fundamental representation of the $SU(2)$ gauge group, 
 and the $SU(2)$ vector multiplet consisting of vector and chiral 
 superfields $(V,A)$ in the adjoint representation. 
There is a global symmetry $SO(4)\times U(1)_B$, 
 where $SO(4)$ is a quark flavor symmetry 
 and $U(1)_B$ is a baryon number symmetry. 
According the prescription in Ref. \cite{murayama}, 
 we incorporate a baryon chemical potential $\mu$ into this theory 
 by introducing a background fictitious $U(1)_B$ super gauge field $V_B$ 
 with an appropriate vacuum expectation values (see Eq.~(\ref{vev}) ). 
A chiral superfield $A_B$ which is a superpartner of $V_B$ 
 is also introduced because of ${\cal N}=2$ SUSY. 
The chemical potential leads to a tachyonic mass term for 
 scalar quarks so that they immediately undergoes Bose-Einstein
 condensations and the baryon number symmetry is broken 
 at the classical level. 
Then, to realize a classical vacuum preserving 
 the baryon number symmetry, which we are interested in, 
 we further introduce ${\cal N}=1$ SUSY preserving soft mass term 
 for the adjoint chiral gauge multiplet. 
Putting all together, the resulting classical Lagrangian 
 of our theory is defined as
\begin{eqnarray}
 {\cal L}&=&{\cal L}_{\mbox{\tf HM}}
           +{\cal L}_{\mbox{\tf VM}}
           +{\cal L}_{\mbox{\tf soft}}\,,\label{eq:lag} \\
{\cal L}_{\mbox{\tf HM}}
	&=&\sum_{i=1}^2 {\Bigg [}\int d^4\theta
           \left(Q_i^\dagger 
           e^{2 V + 2 S V_B} Q_i
	   +\tilde{Q}_ie^{-2 V - 2 S V_B}
           \tilde{Q}_i^\dagger \right)  \nonumber \\
        & &\left.
           +\sqrt{2}\left({\int d^2\theta \tilde{Q}_i
           \left(A + S{m \over \sqrt{2}}\right) Q_i +h.c.} 
           \right)\right],\\
{\cal L}_{\mbox{\tf VM}}
        &=&\frac{2}{4\pi}
           \mbox{\rm Im}\left[\mbox{\rm tr}
	   \left\{\tau
	   \left({\int d^4\theta A^\dagger 
            e^{2V}A e^{-2V}}
           +\frac{1}{2}\int d^2\theta W^2
	   \right)\right\}\right]\,, \\
 {\cal L}_{\mbox{\tf soft}}&=&m_{\mbox{\tf ad}}\int d^2\theta~{\rm tr}A^2
           +h.c.\,,
\end{eqnarray}
where $\tau={4\pi i \over g^2}+{\theta \over 2\pi}$, 
 $W$ is a $SU(2)$ gauge superfield strength,
 $S=1$ is a baryon number quark superfield carries, 
 and $m_{\mbox{\tf ad}}$ is the adjoint mass.
Here,  we took the normalization 
 $T(R)\delta^{ab}={\rm tr}(T^aT^b)={1 \over 2}\delta^{ab}$. 
The vacuum expectation values for vector and chiral superfields 
 $(V_B,A_B)$ are taken to be \cite{murayama} 
\begin{eqnarray}
 &\langle V_B \rangle=\theta\sigma^\mu\bar{\theta}\langle A_\mu \rangle
 ~~~\mbox{\rm with}~~~\langle A_\mu \rangle=\left({\mu, 0, 0,
0}\right),& 
\nonumber \\
 &\langle A_B \rangle=0.& \label{vev}
\end{eqnarray}
In general one can take the non-zero vacuum expectation value 
 for $A_B$ and introduce it as a soft SUSY breaking term. 
We fix it to be zero, for simplicity. 
 
The scalar potential is written down as
\begin{eqnarray}
 V&=&q_i^\dagger\left(2|A+S m/\sqrt{2}|^2 - \mu^2 \right)q_i
     +\tilde{q}_i\left(2|A+S m/\sqrt{2}|^2 - \mu^2 \right)\tilde{q}_i^\dagger 
     \nonumber \\
  & &+{1 \over 2b}\sum_{a=1}^3
     (q_i^\dagger T^a q_i - \tilde{q}_i T^a \tilde{q}_i^\dagger)^2
     + b {\rm tr}[A,A^\dagger]^2 + q_i^\dagger[A^\dagger,A]q_i
     -\tilde{q}_i[A^\dagger,A]\tilde{q}_i^\dagger \nonumber \\
  & &+{1 \over b}\sum_{a=1}^3|\sqrt{2}q_iT^a\tilde{q}_i
     +m_{\mbox{\tf ad}}A^a|^2,
\end{eqnarray}
where the quantity $b$ is the coupling constant defined by
 $b\equiv {1 \over 4\pi}{\rm Im}\tau$.
The fields
 $A,~q_i$ and $\tilde{q}_i$ are the scalar components of the
 corresponding chiral superfields, respectively.
The index $i$ for the flavor symmetry is implicitly summed. 

\begin{figure}[t]
\begin{center}
\leavevmode
  \epsfysize=5.0cm
  \epsfbox{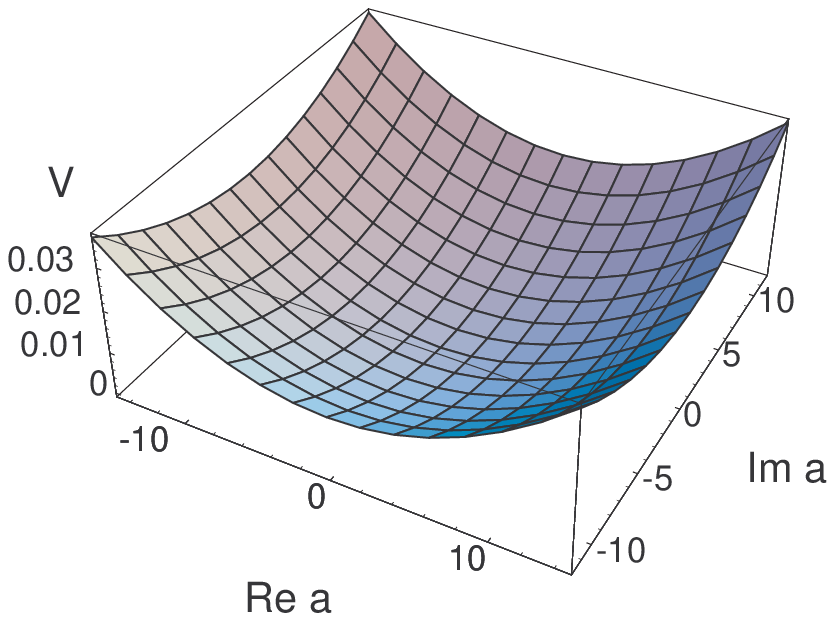} \\ 
\caption{Potential as a function of 
 $a$ for $b=1$, $\mu=0.06$, $m_{\mbox{\tf ad}}=0.01$ and $m=5$.}
\label{fig2}
\end{center}
\begin{center}
\leavevmode
  \epsfysize=5.0cm
  \epsfbox{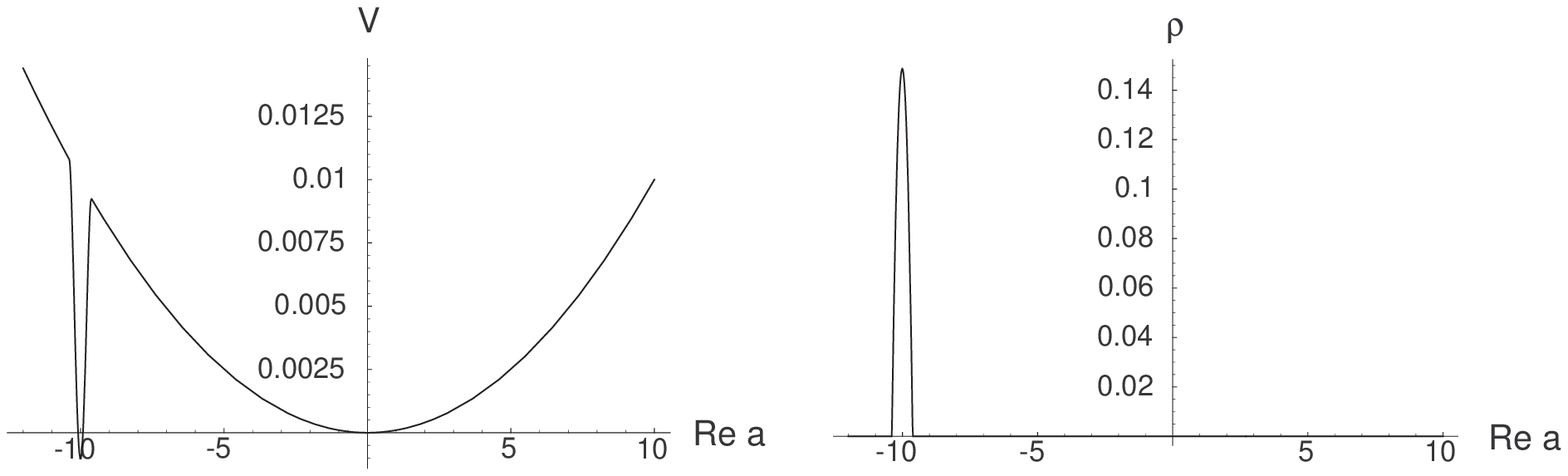} \\ 
\caption{Plots of potential and quark condensation 
 along real $a$ axis
 for $b=1$, $\mu=0.06$, $m_{\mbox{\tf ad}}=0.01$ and $m=5$.}
\label{fig3}
\end{center}
\end{figure}
Solving stationary conditions with respect to the hypermultiplet
 scalars $q_i$ and $\tilde{q}_i$ yields 
\footnote{
 In the case with $\mu > m$, 
 we also have another solution 
 $|v_i|^2=2b(\mu^2-m^2)\left(1+\sqrt{1-{4bm^2 \over m_{\tf ad}^2}}\right)$
 and
 $|\tilde{v}_i|^2=2b(\mu^2-m^2)\left(1-\sqrt{1-{4bm^2 \over
 m_{\tf ad}^2}}\right)$. 
 This solution breaks the baryon number and out of our interests. 
 In addition, we will consider the opposite case $\mu < m$ in the
 following. 
}
\begin{eqnarray}
 &q_i=\left(\begin{array}{c}
            v_i \\
            0
	   \end{array}
     \right),~~~
  \tilde{q}_i=\left(\begin{array}{c}
            \tilde{v}_i \\
            0
	   \end{array}
     \right)\,,~~~~
v_i=\rho_ie^{i\alpha}\,,~~~\tilde{v}_i=\rho_i\,, & \label{repara} \\
&A={1 \over 2}\left(
\begin{array}{cc}
 a & 0 \\
 0 & -a
\end{array}
\right)\,,~~~a\in {\bf C}\, ,&
\end{eqnarray}
from which we find 
\begin{eqnarray}
 V=-{\rho_i^4 \over 2b}+{|m_{\mbox{\tf ad}}a|^2 \over b}\,,\label{potential}
\end{eqnarray}
with the quark condensation  
\begin{eqnarray}
 \rho_i^2=\left\{
  \begin{array}{l}
   2b\left(\mu^2-2\left|{1 \over 2}a+{m \over \sqrt{2}}\right|^2\right)
   +\sqrt{2}|m_{\mbox{\tf ad}}a| \; \; (\mbox{for } \rho_i^2 >0) \,, \\
   0 \; \; (\mbox{otherwise})\,.
  \end{array}
 \right. \label{condensate}
\end{eqnarray}
The vacuum expectation value of the condensation 
 $\langle \rho \rangle$ is an order parameter 
 for the baryon number symmetry breaking. 

We found that the theory has a vacuum (a local minimum) 
 where the baryon number symmetry is preserved.
Figs.~\ref{fig2} and \ref{fig3} show 
 an example realizing such a situation. 
A 3D plot of the effective potential for the complex variable $a$ 
 is depicted in Fig.~\ref{fig2}. 
In the parameter choice shown in the caption in Fig.~\ref{fig2},
 one can see that the theory has two discrete vacua 
 along the real $a$ axis.
The plots in Fig.~\ref{fig3} show the potential (left) 
 and the quark condensation (right) along real $a$ axis. 
At a vacuum with $\langle a \rangle \neq 0 $, 
 the $SU(2)$ gauge symmetry is broken down to $U(1)$ 
 and quarks become massless (quark singular point). 
Since the quarks condensate there, 
 the baryon number symmetry is broken. 
On the other hand, 
 we find a local minimum at the origin, 
 where there is no quark condensation and 
 the baryon number symmetry is preserved. 
This is the classical vacuum which we are interested in. 

\begin{figure}
\begin{center}
\leavevmode
  \epsfysize=5.0cm
  \epsfbox{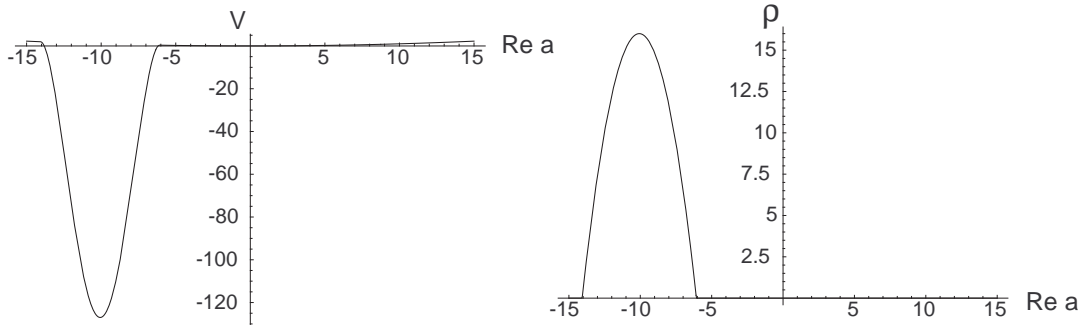} \\ 
\caption{Plots of potential and quark condensation 
 along real $a$ axis for $b=1$, $\mu=2.7$, 
 $m_{\mbox{\tf ad}}=0.1$ and $m=5$.}
\label{fig4}
\end{center}
\end{figure}

We also investigate the behavior of the scalar potential 
 with respect to other choices of the parameters. 
First of all, the quark singular point depends only on $m$. 
As $\mu$ decreases with $m$ and $m_{\mbox{\tf ad}}$ fixed,
 the quark condensation becomes small and the vacuum energy 
 at the quark singular point increases. 
For vanishing $\mu$, ${\cal N}=1$ SUSY is restored  
 and  two equivalent discrete vacua appear. 
As $\mu$ increases, the quark condensation grows 
 and its width becomes wider and reaches the origin. 
As a result, the minimum at the origin is smeared away 
 and the baryon number preserving vacuum disappears 
 (see Fig.~\ref{fig4}). 
As $m$ decreases with $\mu$ and $m_{\mbox{\tf ad}}$ fixed, 
 the quark singular point approaches to the origin
 and, eventually, 
 the quark condensation covers the origin and eliminates 
 the minimum at the origin. 
As $m_{\mbox{\tf ad}}$ decreases with $\mu$ and $m$ fixed, 
 the potential becomes flatter with the quark condensation 
 being small. 
In the limit of vanishing $m_{\mbox{\tf ad}}$ 
 the flat direction appears except around quark singular point 
 at which vacuum energy is negative due to the quark condensation 
 caused by non-zero $\mu$. 

As discussed above, the vacuum preserving 
 the baryon number symmetry 
 is smeared away for small $m$ and large $\mu$. 
From Eq.~(\ref{condensate}), it is easy to find 
 that a vacuum preserving the baryon symmetry exist 
 if the condition $\mu < m$ is satisfied. 
Here, we assume real $m$, for simplicity. 
This situation is similar to 
 the case of ${\cal N}=1$ SUSY QCD investigated in
 Ref. \cite{murayama}, 
 where the authors introduced soft SUSY breaking masses, 
 corresponding to $m$ in our analysis, 
 to provide a classical vacuum preserving 
 the baryon number symmetry. 
In our ${\cal N}=2$ gauge theory, 
 we need to introduce the adjoint mass term 
 as understood from the above discussions. 

\section{Quantum theory}
In this section, we describe low energy Wilsonian effective
 Lagrangian of our theory.
Firstly we recall the result without the chemical potential.
In this case, effective action is just of ${\cal N}=2$ SUSY QCD
 perturbed by adjoint mass term  preserving ${\cal N}=1$ SUSY. 
For the small SUSY breaking parameters $m_{\mbox{\tf ad}}\ll\Lambda$, 
 the effective action is described as \cite{witten} 
\begin{eqnarray}
 {\cal L}_{\mbox{\tf eff}}&=&{\cal L}_{\mbox{\tf SQCD}}
            +{\cal L}_{\mbox{\tf soft}}\,, \\
 {\cal L}_{\mbox{\tf soft}}&=& m_{\mbox{\tf ad}}\int d^2 \theta U(A) + h.c.\,,
\end{eqnarray}
where ${\cal L}_{\mbox{\tf SQCD}}$ is the effective Lagrangian 
 containing full SUSY quantum corrections and
 $U(A)$ is a superfield whose scalar
 component is $u(a)={\rm tr}A^2$.
The first term consists of vector multiplet and
 hypermultiplet parts as will be seen below.
The vector multiplet part is described by prepotential being a function
 of a chiral superfield $A$, vector superfield $V$ and their dual fields
 $A_D$ and $V_D$.
The hypermultiplet part is described by a light hypermultiplet
 with appropriate quantum numbers $(n_e,n_m)_S$, 
 where $n_e$ is electric charge, $n_m$ is magnetic charge, 
 and $S$ is $U(1)_B$ charge. 
They couple to vector multiplet superfields $V(V_D)$ and $A(A_D)$ 
 according to their quantum numbers. 
In the effective Lagrangian around a singular point, 
 the hypermultiplet is expected to be light 
 and enjoys correct degrees of freedom in the effective theory. 

Next we incorporate the baryon chemical potential into this system.
As in the same manner in the classical theory, 
 the fictitious $U(1)_B$ gauge multiplet is introduced  
 and vacuum expectation values for $V_B$ and $A_B$ are 
 taken to be the same as in the classical theory (see Eq.~(\ref{vev})). 
The light hypermultiplet couples to the gauge multiplet 
 according to its $U(1)_B$ charge. 
One might consider to introduce a dual field of $U(1)_B$ gauge 
 multiplet in the strong coupling region of the moduli space.
However, this is not the case, 
 because the $U(1)_B$ gauge multiplet is not a dynamical field 
 and does not affect the $SU(2)$ gauge dynamics. 
Resulting effective action is described by
\begin{eqnarray}
 {\cal L}_{\mbox{\tf eff}}
&=&{\cal L}_{\mbox{\tf HM}}+{\cal L}_{\mbox{\tf VM}}
+{\cal L}_{\mbox{\tf soft}}\,, \label{final-lag}\\
{\cal L}_{\mbox{\tf HM}}
   &=&\sum_{i=2}^{N_f}\left[\int d^4\theta
              \left(H_i^\dagger 
              e^{2n_m V_{D}+2n_eV+2SV_B}H_i
                 +\tilde{H_i}
                 e^{-2n_m V_{D}-2n_eV-2SV_B}
                 \tilde{H}_i^\dagger
              \right)
            \right.
        \\ \nonumber
   & & +\left.
        \left(\int d^2\theta \tilde{H}_i(n_mA_{D}+n_eA +Sm/\sqrt{2})H_i
               +h.c.\right)\right]\,, \\
{\cal L}_{\mbox{\tf VM}}
        &=&\frac{1}{4\pi}
           \mbox{\rm Im}
	   \left(\int d^4\theta
           \frac{\partial F(A)}{\partial A}
           A^\dagger
         + \int d^2\theta \frac{1}{2} 
           \tau W^2
           \right)\,.
\end{eqnarray} 
Here $H_i$ and $\tilde{H}_i$
 denote light quarks, light monopoles or light dyons, 
 and the function $F(A)$ is a prepotential. 
The effective coupling $\tau$ is written by the prepotential as
\begin{eqnarray}
\tau=\frac{\partial^2 F(a)}{\partial a^2}\,.\label{eq:coupling}
\end{eqnarray}

The scalar potential is easily read off from the Lagrangian (\ref{final-lag}) 
 as
\begin{eqnarray}
 V&=&\left(2|a+ S m/\sqrt{2}|^2-\mu^2\right)(|h_i|^2+|\tilde{h}_i|^2)
     +{1 \over 2b}(|h_i|^2-|\tilde{h}_i|^2)^2 \nonumber \\
  & &+{1 \over b}(2|h_i\tilde{h}_i|+|m_{\mbox{\tf ad}}\kappa|^2)
     +{2\sqrt{2} \over b}{\rm Re}(h_i\tilde{h}_i\bar{m}_{\mbox{\tf ad}}
      \bar{\kappa})\,,
\end{eqnarray}
where $a$, $h_i$ and $\tilde{h}_i$ are scalar fields of corresponding
 superfields, and
 $\kappa={\partial u \over \partial a}$.
Solving stationary conditions with respect to light hypermultiplets, 
 we find
 \footnote{Similarly to the classical case, we also have another solution
 which does not exist in the case $\mu>m$. 
 In what follows, we will not consider this solution, 
 since we consider the solution of ${\cal N}=2$ SUSY QCD perturbed 
 by a small $\mu$.}
\begin{eqnarray}
 V=-{2 \rho_i^4 \over b}+{|m_{\mbox{\tf ad}}\kappa|^2 \over b}\,, \label{pot-q}
\end{eqnarray}
where $\rho_i$ is the vacuum expectation value of the condensation of
 light matters
\begin{eqnarray}
 \rho_i^2=\left\{
\begin{array}{l}
 {b \over 2}\left(\mu^2-2\left|a+S{m \over \sqrt{2}}\right|^2\right)
 +{1 \over \sqrt{2}}|m_{\mbox{\tf ad}}\kappa| 
 \,\,  (\mbox{for } \rho_i^2 >0)\,,  \\ 
 0 \, \,  (\mbox{otherwise})\,.  
\end{array}
\right. \label{cond-q}
\end{eqnarray}
Here, we took the phase $m_{\mbox{\tf ad}}\kappa
 =-e^{i\alpha}|m_{\mbox{\tf ad}}\kappa|$ so that the
 potential is minimized, for simplicity.

We supposed that the potential is described by the adequate variables
 associated with the light BPS states. 
For instance, the variable $a$ is understood implicitly as $-a_{D}$, 
 when we consider the effective potential for the monopole.
The same assumption is implicitly included in the following analysis.

\section{Numerical analysis on effective potential}

In the following, we investigate the potential minimum 
 on the $u$-plane numerically. 
The scalar potential is a function of $a(u)$, $a_D(u)$,
 gauge coupling $b={1 \over 4\pi}{\rm Im}\tau$ and its dual coupling. 
All their explicit forms needed for our analysis are given 
 in Ref.~\cite{arai}.
Here, we briefly summarize the results
 for readers' convenience.
 
The scalar fields $a$ and  $a_{D}$ are obtained as periods of 
 the elliptic curves.
The elliptic curves of ${\cal N}=2$ SUSY QCD 
 with $N_f=2$ hypermultiplets having the same mass $m$ 
 were found to be \cite{witten} 
\begin{eqnarray}
y^2=x^2(x-u)+P_{N_f}(x,u,m, \Lambda)\,,
 \label{eq:curve}
\end{eqnarray}
where $\Lambda$ is a dynamical scale of the $SU(2)$ gauge interaction.
In this case, the polynomials $P_{N_f}$ is given by
\begin{eqnarray}
P_{N_f=2}&=&-\frac{\Lambda^4}{64}(x-u)
       +\frac{\Lambda^2}{4}
       m^2x-\frac{\Lambda^4}{32}m^2.
\end{eqnarray}
The mass formula of the BPS state 
 with the quantum number $(n_e, n_m)_S$ 
 is given by $M_{BPS}=\sqrt{2}| n_m a_{D}-n_e a+S m/\sqrt{2}|$. 
If $\lambda$ is a meromorphic differential on the 
curve Eq.~(\ref{eq:curve}) such that
\begin{eqnarray}
\frac{\partial \lambda}{\partial u}
       =\frac{\sqrt{2}}{8\pi}\frac{dx}{y},
\end{eqnarray}
the periods are given by the contour integrals
\begin{eqnarray}
  a_{D}=\oint_{\alpha_1}\lambda, \; 
  a=\oint_{\alpha_2}\lambda,
 \label{eq:period}
\end{eqnarray}
where the cycles $\alpha_1$ and $\alpha_2$ are defined 
 so as to encircle $e_2$ and $e_3$, and $e_1$ and $e_3$, respectively,
 where they are roots of the algebraic curve Eq. (\ref{eq:curve})
 in Weierstrass normal form (see Eqs. (\ref{weierstrass}) and (\ref{eq:root})).
Meromorphic differentials are given by  
\begin{eqnarray}
\lambda_{SW}^{(N_f=2)}
     &=&-\frac{\sqrt{2}}{4\pi}
        \frac{ydx}{x^2-\frac{\Lambda^4}{64}} 
      =-\frac{\sqrt{2}}{4\pi}\frac{dx}{y}
        \left[ x-u  +\frac{m^2\Lambda^2} 
         {4\left(x+\frac{\Lambda^2}{8} \right)}\right] \, .
     \label{eq:f2}
\end{eqnarray}
They have a single pole at $x=-\frac{\Lambda}{8}$ 
 and the residue is given by 
\begin{eqnarray}
\mbox{\rm Res}\lambda_{SW}^{(N_f)}
   =\frac{1}{2\pi i}(-1)\frac{m}{\sqrt{2}}.
\end{eqnarray} 

We calculate the periods by using the Weierstrass normal form 
 for later convenience. 
In this form, the algebraic curve is rewritten 
 by new variables $x=4X+\frac{u}{3}$ and $y=4Y$, such that 
\begin{eqnarray}
Y^2 = 4X^3-g_2X-g_3
   &=&4(X-e_1)(X-e_2)(X-e_3),\label{weierstrass}\\ \nonumber
\sum_{i=1}^3e_i&=&0,
\end{eqnarray}
where $g_2$ and $g_3$ are explicitly written by 
\begin{eqnarray}
g_2&=&\frac{1}{16}\left(\frac{4}{3}u^2
                   +\frac{\Lambda^4}{16}-m^2 
                    \Lambda^2\right),\\
g_3&=&\frac{1}{16}\left(
                   \frac{m^2 \Lambda^4}{32}
                   -\frac{u}{12}m^2\Lambda^2
                   -\frac{u\Lambda^4}{96}
                   +\frac{2u^3}{27}\right).
\end{eqnarray}
Converting the Seiberg-Witten differentials of 
 Eq.~(\ref{eq:f2}) 
 into the Weierstrass normal form 
 and substituting it into Eq.~(\ref{eq:period}), 
 we obtain the integral representations of the periods as follows 
 ($a_{D}$ and $a$ are denoted by $a_{1}$ and $a_{2}$, respectively):
\begin{eqnarray}
 a_{i}&=&-\frac{\sqrt{2}}{4\pi}
    \left(-\frac{4}{3}uI_1^{(i)}+8I_2^{(i)}
    +\frac{m^2\Lambda^2}{8}
    I_3^{(i)}
    \left(c \right)\right), \end{eqnarray}
where $c=-\frac{u}{12}-\frac{\Lambda^2}{32}$
 is the pole of the differentials.
Integrals $I_1^{(i)},I_2^{(i)}$ and $I_3^{(i)}$ are defined as 
\begin{eqnarray}
I_1^{(i)}=\frac{1}{2}\oint_{\alpha_i}\frac{dX}{Y}, \; \; 
I_2^{(i)}=\frac{1}{2}\oint_{\alpha_i}\frac{XdX}{Y}, \; \; 
I_3^{(i)}(c)=\frac{1}{2}
            \oint_{\alpha_i}\frac{dX}{Y(X-c)}.
\end{eqnarray}
The roots $e_i$ of the polynomial defining the cubic 
 are chosen so as to lead to the correct asymptotic behavior 
 for large $|u|$, 
\begin{eqnarray}
a_{D}(u) 
 \sim i\frac{4-N_f}{2\pi}\sqrt{2u} \log\frac{u}{\Lambda^2}, \; \; 
a(u)\sim\frac{\sqrt{2u}}{2}.
\end{eqnarray}
A correct choice is the following: 
\begin{eqnarray}
e_1 &=&\frac{u}{24}-\frac{\Lambda^2}{64}
     -\frac{1}{8}\sqrt{u+\frac{\Lambda^2}{8}
                  +\Lambda m}
                 \sqrt{u+\frac{\Lambda^2}{8}
                  -\Lambda m},
                 \label{eq:root} \\
e_2 &=&\frac{u}{24}-\frac{\Lambda^2}{64}
     +\frac{1}{8}\sqrt{u+\frac{\Lambda^2}{8} + \Lambda m}
                 \sqrt{u+\frac{\Lambda^2}{8} - \Lambda m},
                 \nonumber \\
e_3 &=&-\frac{u}{12}+\frac{\Lambda^2}{32}. 
                 \nonumber
\end{eqnarray}
Fixing the contours of the cycles relative to the positions 
 of the poles, which is equivalent to fix the $U(1)$ 
 baryon numbers for the BPS states, 
 the final formulae are given by 
\begin{eqnarray}
a_{i}&=&-\frac{\sqrt{2}}{4\pi}
    \left(-\frac{4}{3}uI_1^{(i)}+8I_2^{(i)}
    +\frac{m^2\Lambda^2}{8}
    I_3^{(i)}
    \left(-\frac{u}{12}
    -\frac{\Lambda^2}{32}\right)\right)
    -\frac{m}{\sqrt{2}}\delta_{i2},
    \label{eq:period2} 
\end{eqnarray}
with the integral $I_j^{(1)} \; (j=1,2,3)$ explicitly given by
\begin{eqnarray}
I_1^{(1)}&=&\int_{e_2}^{e_3}\frac{dX}{Y}
          = \frac{iK(k^\prime)}{\sqrt{e_2-e_1}}, 
            \label{eq:formula1} \\
I_2^{(1)}&=&\int_{e_2}^{e_3}\frac{XdX}{Y}
          = \frac{ie_1}{\sqrt{e_2-e_1}}K(k^\prime)
            +i\sqrt{e_2-e_1}E(k^\prime), 
            \label{eq:formula2} \\
I_3^{(1)}&=&\int_{e_2}^{e_3}\frac{dX}{Y(X-c)}
             = \frac{-i}{(e_2-e_1)^{3/2}}
            \left\{
            \frac{1}{k+\tilde{c}}K(k^\prime)
            +\frac{4k}{1+k}
            \frac{1}{\tilde{c}^2-k^{2}}
            \Pi_1\left(\nu,\frac{1-k}{1+k}         
                \right)            
            \right\},
            \label{eq:formula3}
\end{eqnarray}
where $k^2 = \frac{e_3-e_1}{e_2-e_1}$, 
      $k^{\prime 2}=1-k^2=\frac{e_2-e_3}{e_2-e_1}$,  
      $\tilde{c}= \frac{c-e_1}{e_2-e_1}$, 
and $\nu=-\left(\frac{k+\tilde{c}}{k-\tilde{c}}\right)^2
          \left(\frac{1-k}{1+k}\right)^2$. 
The formulae for $I_j^{(2)}$ are obtained from $I_j^{(1)}$ 
 by exchanging the roots, $e_1$ and $e_2$. 
In Eqs.~(\ref{eq:formula1})-(\ref{eq:formula3}), 
 $K$, $E$, and $\Pi_1$ are the complete elliptic integrals \cite{higher} 
 given by
\begin{eqnarray}
K(k)&=&\int_0^1 \frac{dx}
     {\left[(1-x^2)(1-k^2x^2)\right]^{1/2}}, 
       \\ \nonumber
E(k)&=&\int_0^1 dx\left(\frac{1-k^2x^2}{1-x^2}
     \right)^{1/2}, 
       \\ \nonumber
\Pi_1(\nu,k)&=&\int_0^1\frac{dx}
     {[(1-x^2)(1-k^2x^2)]^{1/2}(1+\nu x^2)}.
\end{eqnarray}
The effective coupling $\tau$ and $\kappa$ can be calculated
 by using above results.
\begin{eqnarray}
 \tau&=&\frac{\partial a_{D}}{\partial a}=\frac{\omega_1}{\omega_2}\,, \\
  \kappa&=&{\partial u \over \partial a}={1 \over \omega_{1}}\,,
\end{eqnarray}
where $\omega_i$ is the period of the Abelian differential, 
\begin{eqnarray}
\omega_i=\oint_{\alpha_i}\frac{dX}{Y}=2I_1^{(i)} \; ~(i=1,2)\,.
\end{eqnarray}

Before going to numerical analysis of the effective potential, 
 let us see behaviors of the singular point 
 with respect to the mass of hypermultiplets. 
The effective potential is expected to have its minimum  
 around the singular points, 
 because the singular point is energetically favored 
 due to the non-zero condensation (see Eq.~(\ref{pot-q})) 
 of the light BPS state such as a quark, monopole or dyon
 with appropriate quantum number $(n_e,n_m)_S$. 
The solution of the cubic polynomial 
 determines the position of the singular points 
 on the $u$-plane \cite{witten}. 
In $N_f=2$ case with the same masses, the solution is given by 
\begin{eqnarray}
u_1=-m\Lambda-\frac{\Lambda^2}{8}\,, \; 
u_2=m\Lambda-\frac{\Lambda^2}{8}\,, \;
u_3=m^2+\frac{\Lambda^2}{8}\,. \label{descriminat}  
\end{eqnarray}
In the following, we fix the dynamical scale $\Lambda$ 
 to be $\Lambda=2\sqrt{2}$. 
\begin{figure}
\begin{center}
\leavevmode
  \epsfysize=4.7cm
  \epsfbox{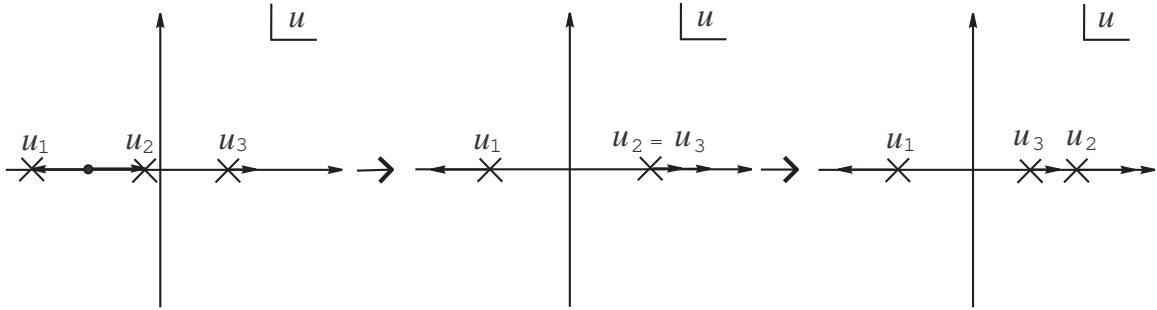} \\ 
\caption{Flow of the singular point in $N_f=2$ case.}
\label{evo2}
\end{center}
\end{figure}
The flow of the singular points with respect to 
 the hypermultiplet mass is sketched in Fig.~\ref{evo2}. 
For $m=0$, the singular points appear 
 at $u_1=u_2=-1$ and $u_3=1$. 
Here, at $u=-1$, two singular points degenerate. 
For non-zero $m>0 $, 
\footnote{ 
 We consider only the case $m > 0$, 
 since the result for $m < 0$ can be obtained 
 by exchanging $u_1 \leftrightarrow u_2$,
 as can be seen from the first two equations in Eq.~(\ref{descriminat}).} 
 this singular point splits into two singular points $u_1$ and $u_2$,
 which correspond to the BPS states with quantum numbers 
 $(1,1)_{-1}$ and $(1,1)_1$, respectively.
As $m$ is increasing, these singular points, $u_1$ and $u_2$, 
 are moving to the left and the right on real $u$-axis, respectively.  
Two singular points, $u_2$ and $u_3$, collide and degenerate 
 at the Argyres-Douglas point \cite{argyres} ($u=\frac{3\Lambda^2}{8}$) for 
 $m= \frac{\Lambda}{2}$, where it is believed that the theory becomes 
 superconformal one.
As $m$ is increasing further, there appear two singular points 
 $u_2$ and $u_3$ again, 
 and quantum numbers of the corresponding BPS states, 
 $(1,1)_1$ and $(0,1)_0$, change into $(1,0)_1$ and $(1,-1)_1$, 
 respectively. 
The singular point $u_2$ is moving to the right faster 
 than $u_3$. 
\begin{figure}[t]
\begin{center}
\leavevmode
  \epsfysize=5.0cm
  \epsfbox{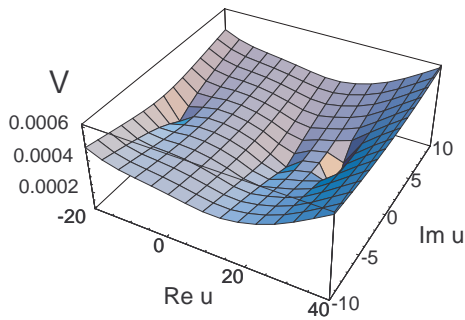} \\ 
\caption{Plot of the potential as a function of $u$
 for $\mu=0.06$, $m_{\mbox{\tf ad}}=0.01$ and $m=5$.}
\label{quant3d}
\end{center}
\end{figure}
\begin{figure}
\begin{center}
\leavevmode
  \epsfysize=5.0cm
  \epsfbox{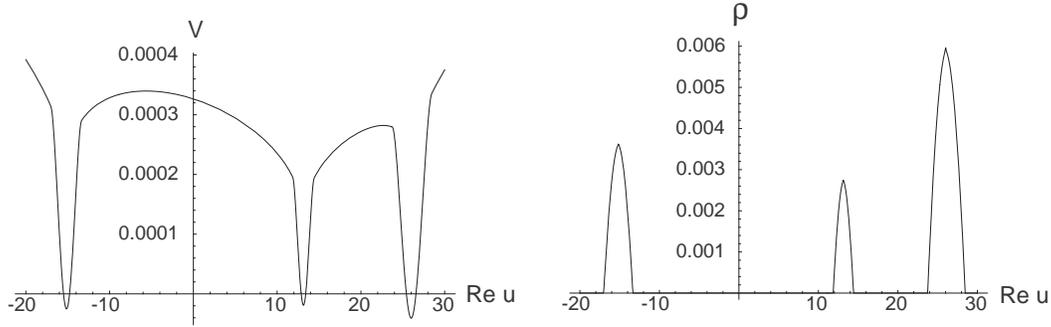} \\ 
\caption{Plots of the potential and the condensations
 for two dyons and quark along real $u$ axis 
 for $\mu=0.06$, $m_{\mbox{\tf ad}}=0.01$ and $m=5$.}
\label{qd2}
\end{center}
\end{figure}

In the following, 
 to see the effect of the dynamics on breaking 
 of the baryon number symmetry, 
 we analyze only the case with $m > \Lambda/2$. 
In this parameter region, 
 the quantum theory has the quark singular point 
 at weak coupling region, 
 which corresponds to the one in the classical theory. 
Therefore, other potential minima, if we find them, 
 correspond to the vacuum preserving the baryon number symmetry 
 at the classical level. 
On the other hand, for $ m < \Lambda/2$, 
 the quantum theory has no quark singular point, 
 while the classical theory has it. 
The quark cannot appear as a light state and, instead, 
 other solitonic state such as dyon and monopole appear 
 due to the strong gauge dynamics at the region $u < \Lambda^2$. 

Now we are ready for numerical analysis. 
The effective potential and the condensations for light matters 
 along real $u$ axis are shown in Figs.~\ref{quant3d} and \ref{qd2}.
When $m$ is real, the singular point is on the real $u$ axis 
 and the effective potential has its minimum there. 
In the left figure in Fig.~\ref{qd2}, 
 the right minimum corresponds to the quark singularity. 
The quark condensates around this singular point 
 is depicted in the right figure in Fig.~\ref{qd2}. 
Since the quark has the $U(1)_B$ baryon number $S=1$, 
 the baryon number symmetry is broken. 
This potential minimum corresponds to 
 the global minimum at the classical theory 
 and the baryon number symmetry is already broken there. 
The left and middle minima in the left figure in Fig.~\ref{qd2}
 correspond to the dyon singular points 
 where they have non-zero vacuum expectation values. 
Since these dyons have $U(1)_B$ baryon number $S=-1$ and $S=1$ 
 for the left and middle singular point, respectively, 
 the baryon number symmetry is broken at these points.
Recall that these minima corresponds to 
 the classical minimum at the origin. 
Therefore, we see that the classical minimum splits into 
 two dyon singular points and the baron number is spontaneously 
 broken due to the strong gauge dynamics. 
\begin{figure}[t]
\begin{center}
\leavevmode
  \epsfysize=5.0cm
  \epsfbox{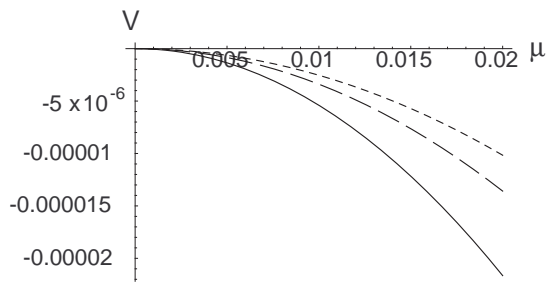} \\ 
\caption{The evolutions of the potential energies at 
 quark (solid line), the left dyon (dashed line) and the middle dyon
 (dotted line) singular points
 as a function of $\mu$ for $m_{\mbox{\tf ad}}=0.01$ and $m=5$.} 
\label{qd3}
\end{center}
\end{figure}
\begin{figure}[t]
\begin{center}
\leavevmode
  \epsfysize=5.0cm
  \epsfbox{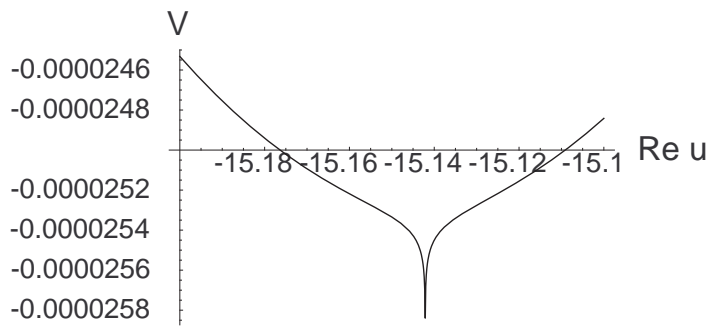} \\ 
\caption{Plot around the left dyon singular point.}
\label{quant2}
\end{center}
\end{figure}

Note that these dyon singular points are not global minimum 
 and the global one is located at quark singular point. 
This situation is similar to the classical theory 
 where the global minimum appears at the quark singular point 
 while the minimum at the origin is the local one. 
Fig.~\ref{qd3} shows the vacuum energy for each singular points 
 as a function of the baryon chemical potential $\mu$.  

The effective potential around the left dyon point 
 is magnified in Fig.~\ref{quant2}. 
We can see the cusp at the minimum, 
 which originates from the non-zero baryon chemical potential. 
The same things happen at other potential minima. 
The existence of the cusps would imply that, 
 for precise description of the effective potential 
 around the potential minima, 
 the periods ($a$ and $a_D$) and the dyon hypermultiplets 
 are not the complete set of dynamical variables and 
 introduction of new variables is necessary, 
 by which the cusps are smoothed away. 
Unfortunately, we could not find such variables. 
However, considering that the cusps disappear 
 in the limit $\mu \rightarrow 0$, 
 we can expect that the role of such variables 
 is just to smooth the cusps away. 
Thus, the vacuum structure of our system would 
 not be drastically changed, and 
 the baryon number is still broken at the minima 
 after such new variables are taken into account.

\begin{figure}[t]
\begin{center}
\leavevmode
  \epsfysize=6.0cm
  \epsfbox{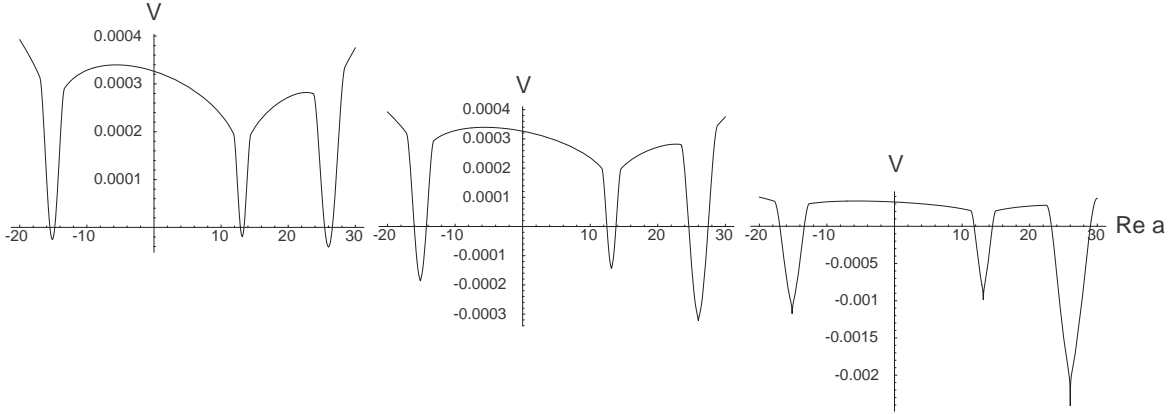} \\ 
\caption{Plots of the scalar potential along real $a$ axis for
 $m_{\mbox{\tf ad}}=0.01$ and $m=5$ with $\mu=0.06$(left), $0.15$(middle), $0.3$(right).}
\label{pert}
\end{center}
\end{figure}

We have found that, once non-perturbative effects are taken into account, 
 the baryon number is always broken for any baryon chemical potential. 
Fig. \ref{pert} shows the effect of the baryon chemical potential 
 on the effective potential. 
We can see that each potential well becomes deeper and wider 
 as the chemical potential is raised (see also Fig. 7). 
When we raise the chemical potential further, 
 potential wells very much overlap with. 
In this case, 
 description of our effective theory by using 
 each suitable dynamical variable around each potential minimum 
 breaks down. 
This corresponds to the breakdown of our effective theory approach 
 for the case with $\mu > \Lambda$. 

\section{Conclusion}

We investigate vacuum structure of ${\cal N}=2$ SUSY QCD 
 in the presence of the baryon chemical potential. 
Our theory is based on the gauge group $SU(2)$ 
 massive $N_f=2$ quark hypermultiplets. 
The baryon chemical potential is introduced 
 in the theory through the fictitious $U(1)_B$ gauge field. 
When the mass term for the adjoint chiral gauge multiplet, 
 which breaks ${\cal N}=2$ SUSY into ${\cal N}=1$, 
 is introduced, 
 the classical theory has two discrete vacua 
 in some parameter region 
 for the chemical potential, the common mass of 
 the quark hypermultiplets and the adjoint mass. 
One is the global minimum where the baryon number is broken, 
 while the other is the local one preserving 
 the baryon number symmetry. 
We concentrated on the local minimum and 
 examined how the baryon number preserving minimum 
 is deformed 
 when all (supersymmetric) quantum corrections are taken into account. 
In our analysis, the SUSY breaking parameters, 
 namely the baryon chemical potential and the adjoint mass, 
 were taken to be much smaller than the dynamical scale  
 of the $SU(2)$ gauge interaction 
 and the exact result on  ${\cal N}=2$ SUSY QCD 
 was used to describe the effective potential 
 at the leading order of perturbation 
 with respect to the small SUSY breaking parameters. 
We found that the classical minimum was deformed 
 due to the strong $SU(2)$ gauge dynamics and 
 the baryon number was spontaneously broken 
 by the dyon condensations. 
Therefore, color superconductivity takes place 
 due to the strong gauge dynamics. 
One the other hand, 
 the global minimum at the classical level  
 lies in the perturbative regime 
 and thus remains at the quantum level. 

\noindent {\large \bf Acknowledgements}

We thank to Masud Chaichian for careful reading of our manuscript.
M.A. is supported by the bilateral program of Japan Society 
 for the Promotion of Science and Academy of Finland, 
 ``Scientist Exchanges.'' 
The work of N.O. is partly supported by 
 the Grant-in-Aid for Scientific Research in Japan (\#15740164).


\end{document}